\documentclass[a4paper,fleqn,usenatbib]{mnras}

\usepackage[T1]{fontenc}
\usepackage{ae,aecompl}

\usepackage{graphicx}	
\usepackage{amsmath}	
\usepackage{amssymb}	

\title[Dynamic changes of EED in 3C 279]{Dynamic changes of emitting electron distribution in the jet of 3C 279: signatures of acceleration and cooling}
\author[D. H. Yan et al.]{Dahai Yan$^{1}$\thanks{E--mail: yandahai@ihep.ac.cn}, Jianjian He$^{1,3}$, Jinyuan Liao$^{1}$, Li Zhang$^{2}$\thanks{E--mail: lizhang@ynu.edu.cn}, Shuang-Nan Zhang$^1$\thanks{E--mail: zhangsn@ihep.ac.cn}\\
$^1$Key Laboratory of Particle Astrophysics, Institute of High Energy Physics,
Chinese Academy of Sciences, Beijing 100049, China\\
$^2$Department of Astronomy, Key Laboratory of Astroparticle Physics of Yunnan Province, Yunnan University, Kunming, 650091, China\\
$^3$University of Chinese Academy of Sciences, Beijing 100049, China}

\date{Accepted XXX. Received YYY; in original form ZZZ}

\pubyear{2015}

\begin{document}
\label{firstpage}
\pagerange{\pageref{firstpage}--\pageref{lastpage}}
\maketitle

\begin{abstract}
We study the dynamic changes of electron energy distribution (EED) through systematically analysing the quasi-simultaneous spectral energy distributions (SEDs) of the flat spectrum radio quasar 3C 279 in different states. With Markov chain Monte Carlo (MCMC) technique we model fourteen SEDs of 3C 279 using a leptonic model with a three-parameter log-parabola electron energy distribution (EED). The 14 SEDs can be satisfactorily fitted with the one-zone leptonic model. The observed $\gamma$ rays in 13 states are attributed to Compton scattering of external infrared photons from a surrounding dusty torus.
 The curved $\gamma$-ray spectrum observed during 2-8 April 2014 is well explained by the external Compton of dust radiation.
 It is found that there is a clear positive correlation between the curvature parameter $b$ of the EED and the electron peak energy $\gamma'_{\rm pk}$. No significant correlation between $b$ and the synchrotron peak frequency $\nu_{\rm s}$ is found, due to the varied product of Doppler factor and fluid magnetic field from state to state. We interpret the correlation of $b-\gamma'_{\rm pk}$ in a stochastic acceleration scenario. This positive correlation is in agreement with the prediction in the stage when the balance between acceleration and radiative cooling of the electrons is nearly established in the case of the turbulence spectral index $q=2$.
\end{abstract}


\section{Introduction}
Blazars are active galactic nucleis (AGNs) who aim their jets almost directly at Earth.
The multi-wavelength spectral energy distribution (SED) of a
blazar has a two-bump shape.
The first bump peaks in infrared to X-ray frequencies, and the second bump peaks in MeV to GeV band.
The first bump is interpreted as synchrotron emission radiated by relativistic electrons in the jet.
The second bump can be produced via inverse Compton (IC) scattering of synchrotron photons \citep[i.e., synchrotron-self Compton: SSC; e.g.,][]{Maraschi1992,Tavecchio98,Finke08,Yan14} and external photons \citep[i.e., external Compton: EC; e.g.,][]{Dermer93,Sikora1994,Kang} by the same population of relativistic electrons that produce the synchrotron emission.

Based on their emission lines' features, blazars are classified as BL Lacertae objects (BL Lacs; having weak or no emission lines) and flat spectrum radio quasars (FSRQs; having strong emission lines).
FSRQs are usually the low synchrotron-peaked blazars (i.e., the synchrotron peak frequency $\nu_{\rm s}<10^{14}$ Hz).
There exist low, intermediate, and high synchrotron-peaked (LSP, ISP, and HSP, respectively,
defined by whether $\nu_{\rm s}<10^{14}$ Hz, $10^{14}<\nu_{\rm s}\ (\rm Hz)<10^{15}$, or $\nu_{\rm s}>10^{15}$ Hz) BL Lacs.
$\gamma$ rays from HSP BL Lacs are thought to be produced by SSC, while $\gamma$ rays from FSRQs are usually attributed to EC.
Due to the intense low-energy photons around the jets, high-energy electrons in FSRQs suffer more severe radiative cooling than those in HSP BL Lacs.

Claims of correlations between
model parameters have been made in blazar studies,
for example, the inverse
correlation between apparent synchrotron
luminosity $L_{\rm syn}$ and $\nu_{\rm s}$, the so-called \emph{blazar sequence} \citep{fos98}.
The \emph{blazar sequence} is interpreted in terms of cooling processes \citep{ghi98,ghisellini08,Finke13}.
Recently, \cite{Dermer15} explained the \emph{spectral-index diagrams}, namely the inverse correlation between $\gamma$-ray photon spectral index $\Gamma_{\gamma}$ and $\nu_{\rm s}$ reported in \citet[][]{Abdo10a,Abdo10b} and \citet{2011ApJ...743..171A}, using a near-equipartition leptonic model with a three-parameter log-parabola electron energy distribution (EED) .

The correlations of model parameters may carry information about the acceleration and energy loss of electrons in blazars jets \citep[e.g.,][]{Tramacere}. In X-ray data analysis on an HSP BL Lac (Mrk 501), \cite{Massaro1,Massaro2} found that its X-ray spectrum is described by a log-parabola function rather than a simple power-law. The log-parabola X-ray spectrum indicates that the distribution of the
electrons that radiate the X-ray emission also has a log-parabola shape \citep{Massaro2,Tramacere07,MassaroE}.
Later, an inverse correlation between the synchrotron spectral curvature parameter $b_{\rm syn}$ and the synchrotron peak frequency $\nu_{\rm s}$ in HSP BL Lacs is found by \citet{Massaro6,MassaroE} and \citet{Tramacere07,Tramacere09}. \cite{Tramacere} explained this correlation in the stochastic acceleration model with a change in the energy diffusion process.
The inverse correlation between $b_{\rm syn}$ and $\nu_{\rm s}$ is compatible with an acceleration-dominated scenario, in which radiative cooling is not relevant and the curvature of EED is anti-correlated with the peak energy of EED \citep{Tramacere}.
In other words, the inverse correlation between $b_{\rm syn}$ and $\nu_{\rm s}$ in HSP BL Lacs essentially stems from the inverse correlation between the curvature of EED and the peak energy of EED; meanwhile,  the $b_{\rm syn}$-$\nu_{\rm s}$ anticorrelation also implies that the value of $B'\cdot\delta_{\rm D}$ for several HSP BL Lacs (e.g., Mrk 421 and Mrk 501) keeps $\sim$ constant from state to state.
Note that the curvature parameter $b_{\rm syn}$ of the synchrotron spectrum  is linearly correlated with $b$:
$b_{\rm syn}\thickapprox b/5$ \citep{Paggi}

As shown in \cite{Yan13} and \cite{Zhou14}, fitting high-quality SED with the Markov chain Monte Carlo (MCMC)
technique enables us to derive the confidence intervals of the model parameters, which is very helpful to
build correlations between model
parameters. In a related modeling effort, \citet{Dermer14} developed a new leptonic modeling approach \citep[also see][]{Cerruti}. This new approach adopts a three-parameter log-parabola EED, and uses observed qualities (e.g., the
apparent synchrotron luminosity and the synchrotron peak frequency) to express the physical parameters
(e.g., fluid magnetic field $B'$ and Doppler factor $\delta_{\rm D}$).
Using this modeling approach and the MCMC technique to fit multiwavelength data, one can simultaneously derive the constraints on the observed qualities and the physical parameters \citep{Yan151}.
On the observation side, tens of simultaneous and high-quality SEDs of blazars have been constructed
\citep[e.g.,][]{Hayashida,Hayashida15,Pacciani,42115aa}, which enable us to systematically analyse the SEDs with our method \citep{Yan151}, and to study the correlations between model parameters.

 \cite{Yan151}\footnote{The synchrotron component of FSRQs peaks in infrared frequencies in $\nu F_{\nu}$ distribution. Due to the lack of the infrared observations, it is hard to measure $b_{\rm syn}$ and $\nu_{\rm s}$ directly from the synchrotron component of FSRQs.} found that for 3C 279 as the peak energy of EED increases, both $\nu_{\rm s}$ and $b$ increase, which does not conform with the inverse correlation in HSP BL Lacs.
 On the other hand, \citet{Tramacere} showed that when radiative cooling becomes relevant, the inverse correlation between the curvature of EED and the peak energy of EED will be broken; as EED approaches the equilibrium, $b$ increases or keeps stable, depending on the turbulence spectral index.

It is therefore worth investigating the dynamic changes of EED through systematically analysing the SEDs of 3C 279 in different states. We apply our modelling approach \citep{Yan151} to 14 high-quality SEDs of 3C 279 \citep{Hayashida,Hayashida15,Paliya}.
We derive 95\% confidence intervals for model parameters.
It is found that there is a clear positive correlation between $b$ and the peak energy of EED in 3C 279.

In Section 2, we introduce the emission model. The results are showed in Section 3.
Discussions on electron acceleration and cooling are presented in Section 4.
We use parameters $H_0=71\rm \ km\ s^{-1}\ Mpc^{-3}$, $\Omega_{\rm m}=0.27$, and $\Omega_{\Lambda}=0.73$.

\begin{table*}
\begin{center}
\caption{Dependences of $\delta_{\rm D}$$^a$, $B^\prime$, and $\gamma_{\rm pk}^\prime$ on input parameters \citep{Dermer15}.}
\begin{tabular}{cccccccccccc}
\hline
${\rm }$ & Coef. & $L_{48}$ & $\nu_{14}$ &  $t_4$ & & $\zeta_s$  & $\zeta_e$ & & $f_0$  & $f_1$ & $f_2$\\
$$ &  & $ $   &   &   &   &    & & &  &  \\
\hline
\hline
 &  &  &   &   &   &  &  & & &  & \\
$\delta_{\rm D}$ & 17.5 & $~~3/16$ & $~~1/8$ & $-1/8$ & & $-7/16$ & $~~1/4$ & & $-7/16$ & $-1/4$ & $-1/8$ \\
$B^\prime({\rm G})$ & 5.0 &  $-1/16$  &  $-3/8$ & $-5/8$ &  & $~13/16$ & $-3/4$ & & $~13/16$ &  $~~3/4$ & $~~3/8$ \\
$\gamma_{\rm pk}^\prime$ & 523  &  $-1/16$ & $~~5/8$ & $~~3/8$ & & $-3/16$ & $~~1/4$ & & $-3/16$ & $-1/4$ & $-5/8$ \\
\hline\end{tabular}
\label{table1}
\end{center}
$^a$ So, e.g., $\delta_{\rm D} \cong 17.5 L_{48}^{3/16}(\nu_{14}/f_2t_4)^{1/8} (f_0\zeta_s)^{-7/16}(\zeta_e/f_1)^{1/4}$, etc. In the blob scenario, the geometry factor $f_0$=1/3. The $b$-dependent factors are $f_1 = 10^{-1/4b}$ and $f_2 = 10^{1/b}$.\\
\end{table*}

\section{Emission model}

\begin{figure*}
	   \centering
		\includegraphics[width=500pt,height=520pt]{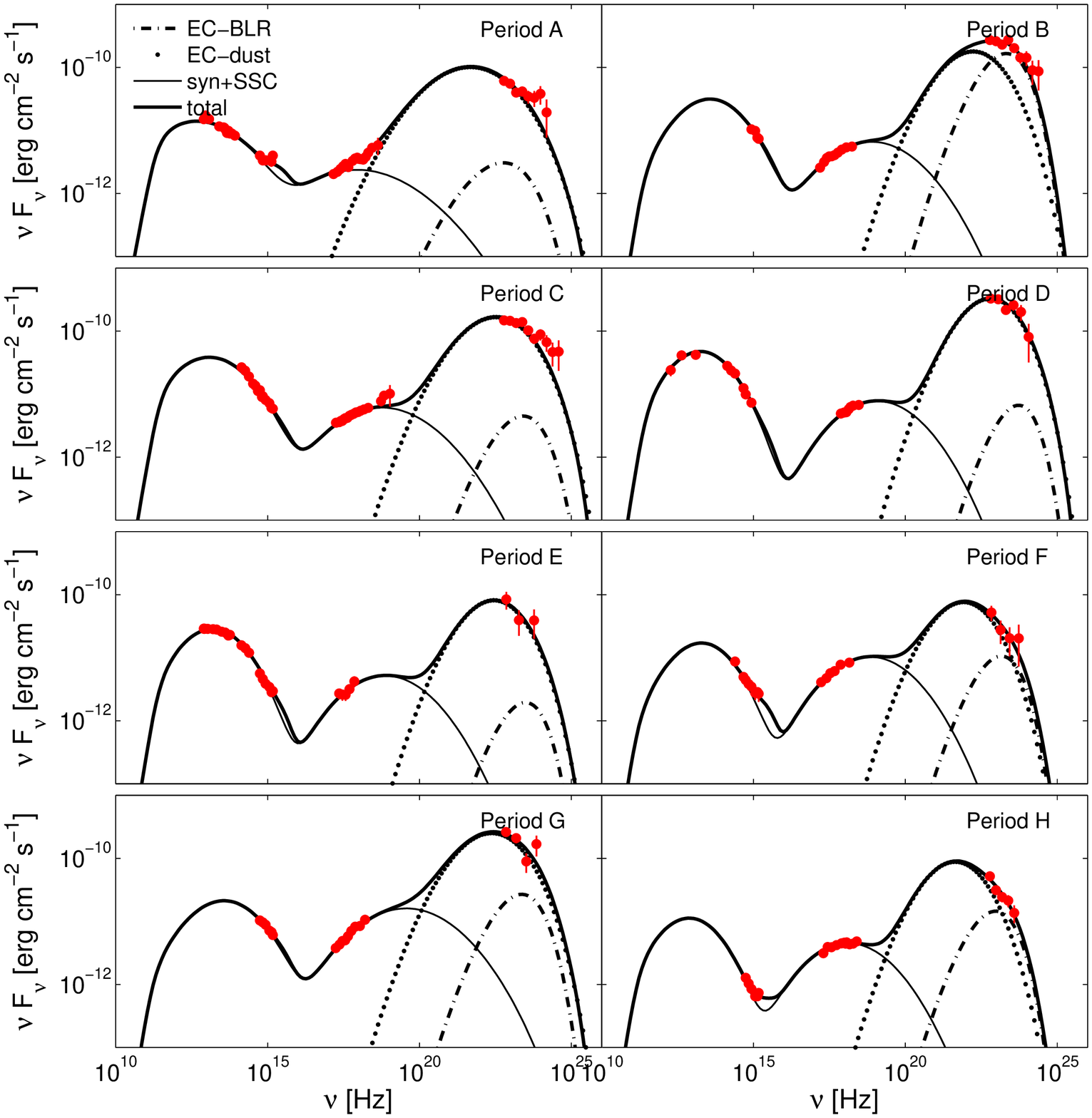}
	  \caption{Best-fitting models to the 8 SEDs of 3C 279 in \citet{Hayashida} . \label{sed1}}
\end{figure*}

\begin{figure*}
	   \centering
		\includegraphics[width=500pt,height=400pt]{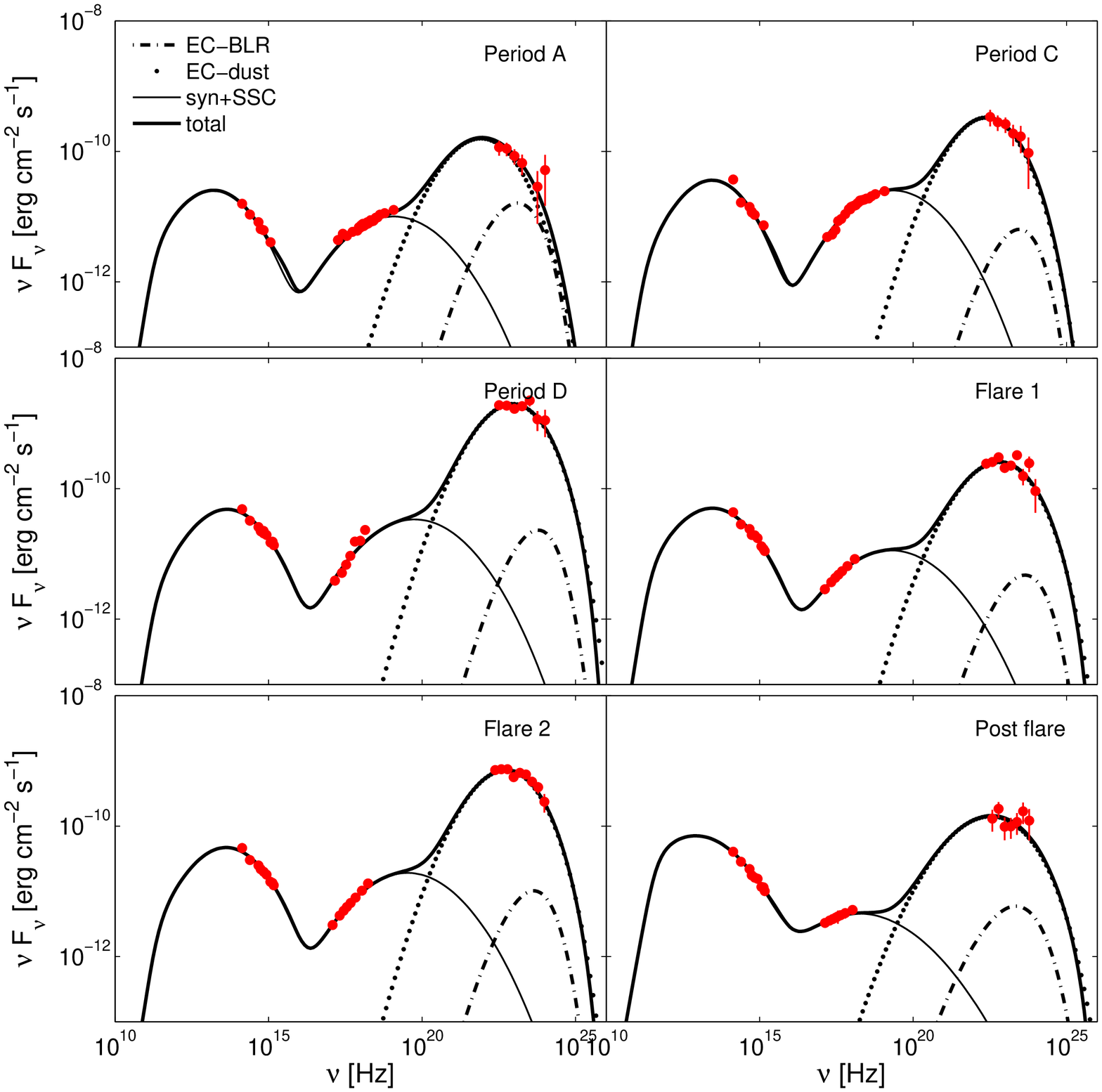}
	  \caption{Best-fitting models to the 3 SEDs of 3C 279 in \citet[][]{Hayashida15} (Periods A, C, D), and to the 3 SEDs in \citet{Paliya} (Flare1, Flare2, Post flare). \label{sed2}}
\end{figure*}

\begin{table*}
\scriptsize
\begin{center}
 \caption{Input parameters$^*$. The mean values and the marginalized 95\% confidence intervals (CI) for free parameters are reported. }
\label{input}
\begin{tabular}{@{}cccccccccccc}
\hline
 & & & & & Input\\
 \hline
&                $\zeta_e$         & $b$        & $L_{48}$    & $\nu_{14}$ & $t_4$        & $\zeta_s$     & $r$                             & $L_{\rm disk}$$^{**}$ & label\\
&                                                 &     &          &            &       &           &(pc)                                                    & ($\rm 10^{46}\ erg\ s^{-1}$)\\
\hline
 & & & & & 2012H\\
\hline
Period A         & 3.26            & 0.68       & 0.11        & 0.13       &  0.77         & 0.25          & 0.25        & 0.23 & a\\
 (95\% CI)       & 1.73-4.86       & 0.61-0.77  & 0.10-0.13   & 0.08-0.21  &  0.35-1.0     & 0.19-0.33     & 0.11-0.56   & -\\
 \hline
Period B         & 1.66            & 1.04       & 0.19        & 0.72       &  0.41         & 0.35          & 0.07        & 0.15 & b\\
 (95\% CI)       & 0.48-2.56       & 0.78-1.44  & 0.15-0.28   & 0.55-0.94  &  0.10-0.94    & 0.18-0.58     & 0.03-0.11   & -\\
\hline
Period C         & 4.65            & 0.95       & 0.28        & 0.25       &  0.89         & 0.20          & 0.88        & 0.15 & c\\
 (95\% CI)       & 3.89-5.0        & 0.88-1.02  & 0.25-0.32   & 0.22-0.28  &  0.67-1.0     & 0.17-0.23     & 0.72-1.04   & -\\
 \hline
Period D         & 3.98            & 1.52       & 0.30        & 0.34       &  0.69         & 0.19          & 0.87         & 0.15 & d\\
 (95\% CI)       & 2.26-5.0        & 1.34-1.73  & 0.27-0.33   & 0.28-0.42  &  0.24-1.0     & 0.15-0.23     & 0.65-1.06    & -\\
 \hline
Period E         & 2.14            & 1.40       & 0.19        & 0.29       &  0.62         & 0.18          & 0.54         & 0.15 & e\\
 (95\% CI)       & 0.41-4.75       & 1.29-1.52  & 0.18-0.20   & 0.26-0.32  &  0.16-1.0     & 0.13-0.24     & 0.11-1.11    & -\\
 \hline
Period F         & 3.31            & 1.57       & 0.11        & 0.39       &  0.61         & 0.73          & 0.19         & 0.25 & f\\
 (95\% CI)       & 0.98-4.93       & 1.20-2.04  & 0.07-0.15   & 0.29-0.60  &  0.16-1.0     & 0.40-1.21     & 0.04-0.62    & -\\
 \hline
Period G         & 4.15            & 1.12       & 0.15        & 0.75       &  0.79         & 0.90          & 0.1          & 0.15 & g\\
 (95\% CI)       & 1.94-5.00       & 1.0-1.25   & 0.13-0.17   & 0.58-1.01  &  0.41-1.0     & 0.65-1.30     & 0.03-0.26    & -\\
 \hline
Period H         & 4.09            & 1.61       & 0.08        & 0.15       &  0.65         & 0.43          & 0.11          & 0.06 & h\\
 (95\% CI)       & 1.91-5.00       & 1.38-1.94  & 0.05-0.12   & 0.11-0.22  &  0.23-1.0     & 0.27-0.68     & 0.04-0.30     & -\\
 \hline
 & & & & & 2015H\\
 \hline
 Period A        & 1.76            & 1.25       & 0.18        & 0.31       &  4.28         & 0.48          & 0.31         & 0.15 & i\\
 (95\% CI)       & 0.85-4.12       & 1.11-1.35  & 0.16-0.19   & 0.26-0.36  &  2.41-5.0     & 0.43-0.53     & 0.12-0.72    & -\\
 \hline
Period C         & 4.1             & 1.50       & 0.23        & 0.55       &  0.94         & 0.82          & 0.23         & 0.15 & j\\
 (95\% CI)       & 2.99-5.00       & 1.43-1.60  & 0.21-0.24   & 0.50-0.61  &  0.63-1.0     & 0.75-0.89     & 0.1-0.41     & -\\
 \hline
Period D         & 12.2            & 1.36       & 0.32        & 0.93       &  0.43        & 0.94           & 0.25         & 0.15 & k\\
 (95\% CI)    & 7.3-15          & 1.15-1.53  & 0.29-0.37   & 0.68-1.12  &  0.25-0.5    & 0.53-1.41      & 0.09-0.52    & -\\
\hline
 & & & & & 2015P\\
 \hline
Flare 1          & 4.23            & 1.15       & 0.35        & 0.59       &  0.39        & 0.27        & 0.74           & 0.15 & l\\
 (95\% CI)       & 2.68-5.0        & 1.01-1.3   & 0.31-0.40   & 0.43-0.75  &  0.19-0.5    & 0.19-0.37   & 0.56-0.92      & -\\
 \hline
Flare 2          & 3.58            & 1.31       & 0.31        & 0.85       &  0.41        & 0.49        & 0.22           & 0.15 & m\\
 (95\% CI)       & 2.24-4.91       & 1.22-1.41  & 0.29-0.34   & 0.74-0.97  &  0.24-0.5    & 0.41-0.58   & 0.1-0.46       & -\\
 \hline
Post flare       & 3.05            & 0.82       & 0.60        & 0.18       &  0.33         & 0.08        & 0.98          & 0.15 & n\\
 (95\% CI)       & 0.75-5.0        & 0.59-1.07  & 0.40-1.09   & 0.04-0.38  &  0.12-0.5     & 0.04-0.14   & 0.5-1.5       & -\\
 \hline
\end{tabular}
\end{center}
$^*$ We fix $T_{\rm BLR}=6.3\times10^4\ $K and $T_{\rm dust}=800\ $K in all fittings.
$^{**}$ There is no evidence for a thermal emission feature in these SEDs, making it difficult to constrain the accretion disk luminosity.
The value of $L_{\rm disk}$ in this table is the maximum disk luminosity allowed by the optical-UV SED, which is fixed in the fitting.
\end{table*}

\begin{table*}
\scriptsize
\begin{center}
 \caption{Output parameters and jet powers. The mean values and the marginalized 95\% confidence intervals (CI) for interested parameters are reported.}
\label{output}
\begin{tabular}{@{}ccccccccc}
 \hline
 & & & & Output\\
 \hline
 & $B'$$^a$ & $\delta_{\rm D}$$^a$  & $\gamma'_{\rm pk}$$^a$  & $R'$$^b$                   &   $u_{\rm dust}$      &  $u_{\rm BLR}$         & $P_{\rm B}$               & $P_{\rm r}$ \\
 & (G) &                  &                    & ($10^{16}\rm \ cm$)    & ($10^{-3}\rm erg\ cm^{-3}$)    & ($10^{-3}\rm erg\ cm^{-3}$)     & ($\rm 10^{45}\ erg\ s^{-1}$)  & ($\rm 10^{46}\ erg\ s^{-1}$)\\
 \hline
 & & & & 2012H\\
 \hline
Period A         & 1.6           & 30      & 50        & 1     &  0.6         & 0.1        & 0.8        & 0.4\\
 (95\% CI)       & 1.0-2.5       & 26-37   & 27-79     & -     &  0.2-0.9     & 0.01-1     & -          & -\\
 \hline
Period B         & 2.5           & 36      & 146        & 0.2    &  0.89         & 5        & 0.6        & 0.4 \\
 (95\% CI)       & 1.5-4.2       & 24-50   & 120-178    & -      &  0.88-0.9     & 1-18     & -          & -\\
\hline
Period C         & 0.6            & 49       & 140        & 1.5    &  0.04         & 0.003           & 0.9    & 0.3\\
 (95\% CI)       & 0.5-0.7        & 44-53    & 129-149    & -      &  0.02-0.08    & 0.002-0.005     & -      & -\\
 \hline
Period D         & 0.5            & 56       & 250        & 1.5    &  0.04         & 0.003           & 0.8     & 0.4\\
 (95\% CI)       & 0.4-0.9        & 47-70    & 173-323    & -      &  0.02-0.1     & 0.002-0.007     & -       & -\\
 \hline
Period E         & 1.3            & 43       & 171        & 1.2    &  0.1          & 0.01          & 0.7       & 0.2\\
 (95\% CI)       & 0.5-3.1        & 30-60    & 102-240    & -      &  0.02-0.9     & 0.001-1       & -         & -\\
 \hline
Period F         & 2.3            & 25       & 200        & 0.5    &  0.7         & 0.3          & 0.4         & 0.5\\
 (95\% CI)       & 1.2-5.2        & 16-35    & 163-239    & -      &  0.1-0.9     & 0.008-15     & -           & -\\
 \hline
Period G         & 1.5           & 26       & 248        & 0.8     &  0.8         & 1          & 0.4          & 1.4\\
 (95\% CI)       & 1.1-2.7       & 20-31    & 200-298    & -       &  0.7-0.9     & 0.1-18     & -            & -\\
 \hline
Period H         & 1.6            & 28       & 143        & 0.6      &  0.8         & 1           & 0.3      & 0.5\\
 (95\% CI)       & 1.0-3.0        & 20-37    & 117-174    & -        &  0.7-0.9     & 0.07-17     & -        & -\\
 \hline
 & & & & & 2015H\\
 \hline
 Period A        & 0.8            & 20       & 266        & 2.7       &  0.4          & 0.06          & 1.8         & 1.9\\
 (95\% CI)       & 0.4-1.2        & 17-26    & 228-304    & -         &  0.08-0.9     & 0.005-1       & -           & -\\
 \hline
Period C         & 1.1             & 28       & 295        & 0.9       &  0.7         & 0.2          & 0.5         & 1.3\\
 (95\% CI)       & 0.9-1.5         & 26-31    & 278-310    & -         &  0.4-0.9     & 0.03-1.5     & -           & -\\
 \hline
Period D         & 0.8              & 44       & 346        & 0.7     &  0.6        & 0.1           & 0.3         & 2.9\\
 (95\% CI)       & 0.5-1.0          & 38-55    & 259- 425   & -       &  0.2-0.9    & 0.01-2.4      & -    & -\\
\hline
 & & & & & 2015P\\
 \hline
Flare 1          & 0.8            & 56       & 204        & 0.8       &  0.07        & 0.005        & 0.6           & 0.3\\
 (95\% CI)       & 0.6-1.2        & 47-66    & 155-255    & -         &  0.03-0.2    & 0.002-0.01   & -      & -\\
 \hline
Flare 2          & 1.3            & 42       & 260        & 0.6       &  0.7        & 0.1        & 0.5           & 1.2\\
 (95\% CI)       & 0.9-1.6        & 37-48    & 229-289    & -         &  0.3-0.9    & 0.02-1.6   & -             & -\\
 \hline
Post flare       & 1.0            & 77       & 66        & 1       &  0.02         & 0.002        & 1.5          & 0.1\\
 (95\% CI)       & 0.5-2.4        & 53-97    & 15-132    & -       &  0.004-0.3     & 0.001-0.02   & -       & -\\
\hline
\end{tabular}
\end{center}
$^a$ derived by the relations showed in Table~\ref{table1}.
$^b$ derived by the relation $R'=t_{\rm min, var}c\delta_{\rm D}/(1+z)$.
\end{table*}

In a one-zone leptonic emission model, multiwavelength emission is assumed to be produced in a blob of comoving radius $R'$ with comoving magnetic field $B'$. This blob moves towards us with relativistic speed, and its bulk Lorentz factor is $\Gamma_{\rm bulk}$. For a blazar, we assume the Doppler factor $\delta_{\rm D}=\Gamma_{\rm bulk}$.

The non-thermal electron distribution is assumed to be isotropic in the blob, and
described by a log-parabola function \citep{Dermer14}
\begin{equation}
\gamma'^2 N'_e(\gamma')\sim\left( \frac{\gamma'}{\gamma'_{\rm pk}} \right)^{-b\ \log{(\gamma'/\gamma'_{\rm pk})}}\;,
\end{equation}
where $\gamma'$ is the electron Lorentz factor; as already noted, $b$ is the curvature parameter of EED. The electron distribution peaks at $\gamma'_{\rm pk}$ in $\gamma'^2 N'_e(\gamma')$ distribution. Because of its continuous curvature, the three-parameter log-parabolic EED produces smooth low-energy cutoff and high-energy cutoff. In our calculations, we integrate $\gamma'_{\rm pk}$ from 1 to $10^{10}$ to avoid arbitrary cutoffs.

Curved EED described by a log-parabola function can be generated in second-order, stochastic acceleration processes \citep[e.g.,][]{Becker06,sp08,Tramacere}.

For FSRQs, low energy photons from its broad-line region (BLR) are important seed photons for IC.
A dilute blackbody is used to describe the BLR radiation \citep{Tavecchio08}.
Given that BLR radiation is dominated by the $\rm Ly\alpha$ line photons, we adopt an effective temperature $T_{\rm BLR}=6.3\times10^4\ $K for the BLR radiation.
We assume that there is a dust torus at larger distance from the central black hole. The IR dust radiation is also assumed to be a dilute blackbody with the temperature of $T_{\rm dust}=800\ $K.

The energy densities of BLR radiation ($u_{\rm BLR}$) and dust radiation ($u_{\rm dust}$) are expressed as the functions of the distance $r$ from the black hole \citep{Sikora09,Hayashida}
\begin{equation}
u_{\rm BLR} (r)=\frac{\tau_{\rm BLR} L_{\rm disk}}{4\pi r^2_{\rm BLR}c[1+(r/r_{\rm BLR})^3]},\
\label{u1}
\end{equation}
\begin{equation}
u_{\rm dust} (r)=\frac{\tau_{\rm dust} L_{\rm disk}}{4\pi r^2_{\rm dust}c[1+(r/r_{\rm dust})^4]}.\
\label{u2}
\end{equation}
The size of BLR is related to the disk luminosity $L_{\rm disk}$: $r_{\rm BLR}=10^{17}(L_{\rm disk}/10^{45}\rm \ erg\ s^{-1})^{1/2}\ $cm \citep{ghisellini09,ghisellini14}. We assume a dust torus with the size of $r_{\rm dust}=10^{18}(L_{\rm disk}/10^{45}\rm \ erg\ s^{-1})^{1/2}\ $cm. Then, we dervie
\begin{equation}
u_{\rm BLR} (r)\simeq\frac{0.3\tau_{\rm BLR}}{1+(r/r_{\rm BLR})^3}\rm \ erg\ cm^{-3},\
\label{u3}
\end{equation}
\begin{equation}
u_{\rm dust} (r)\simeq\frac{0.003\tau_{\rm dust}}{1+(r/r_{\rm dust})^4}\rm \ erg\ cm^{-3}, \
\label{u4}
\end{equation}
where $\tau_{\rm BLR}$ and $\tau_{\rm dust}$ are the fractions of the disk luminosity reprocessed into BLR radiation
and into dust radiation, respectively. The typical values of $\tau_{\rm BLR}=0.1$ \citep[e.g.,][]{ghisellini14} and $\tau_{\rm dust}=0.3$ \citep[e.g.,][]{Hao,Malmrose} are adopted. The energy density of the BLR/dust radiation in the blob comoving frame is $u'_{\rm BLR/dust} (r)=\Gamma^2_{\rm bulk}u_{\rm BLR/dust} (r)$  \citep{ghisellini09}.

Besides the non-thermal emission from the jet, the contribution from the accretion disk is calculated by using the accretion disk luminosity $L_{\rm disk}$.

Besides $T_{\rm BLR}$, $T_{\rm dust}$, and $L_{\rm disk}$, the rest of input parameters in the model are
(i) $t_4=t_{\rm min, var}/[(1+z)10^4\ \textrm{s})]$, the source variability timescale, where $t_{\rm min, var}$ is the minimum variability timescale;
(ii) $L_{48}= L_{\rm syn}/10^{48}~ \textrm{erg s}^{-1}$, the apparent isotropic bolometric synchrotron luminosity;
(iii) the equipartition factor, $\zeta_e={u'_{\rm e}}/{u'_B}$, namely the ratio between the non-thermal electron $(u'_{\rm e})$ and magnetic-field $(u'_B)$ energy densities;
(iv) $\nu_{14}=(1+z){\nu_{\rm s}}/{10^{14}\ \textrm{Hz}}$, the synchrotron peak frequency in the source frame, where $\nu_{\rm s}$ is the synchrotron peak frequency in the observer frame;
(v) $\zeta_s= {u'_{\rm syn}}/{u'_B}$, the ratio between the synchrotron photon $(u'_{\rm syn})$ and magnetic-field energy densities;
(vi) $b$, the curvature parameter of EED, noted above;
and (vii) $r$, the location of emitting blob in the jet.

These input parameters are used to deduce physical model parameters. Table~\ref{table1} shows the dependencies of $B'$, $\delta_{\rm D}$, and $\gamma'_{\rm pk}$ on the input parameters \citep[more details can be found in][]{Dermer14}. We calculate the SSC and EC spectra using the methods given in \citet{Dermer09}. Synchrotron self-absorption (SSA) is included.

The MCMC method is a very powerful fitting tool, which is well suitable to search high-dimensional parameter space, and to evaluate the uncertainties of the model parameters. The details on MCMC technique can be found in \citet{Lewis,yuan11,liu12,Yan13}.

\section{Results}

\citet{Hayashida,Hayashida15} and \citet{Paliya} have constructed 16 high-quality SEDs for 3C 279.
Because there is temporal overlap between Period H in \citet{Hayashida} and the low activity state in \citet{Paliya},
we do not consider the SED in the low activity state in \citet{Paliya}.
Because of the lack of X-ray data in Period B in \citet{Hayashida15}, we do not fit the SED in this period.
Therefore, we totally analyse 14 quati-simultaneous SEDs of 3C 279.

The restriction of the Eddington luminosity on the absolute jet power requires that large
departures from equipartition are not allowed \citep{Dermer15}.
The near-equipartition
log-parabola (NELP) model (with $\zeta_{\rm e}\sim1$) works well for explaining the SEDs of FSRQs \citep {Cerruti,Dermer14,Yan151}.
Here, in order to take into account the impact of the uncertainty of $\zeta_{\rm e}$ on the other parameters,
we allow $\zeta_{\rm e}$ to vary in the range [0.2, 5].
The initial upper limit of $t_4$ in each state is set according to the observations. When there is no information on the variability timescale, we use $t_4=1$, namely the dynamical crossing time associated with the Schwarzschild radius of a $10^9M_{\odot}$ black hole, as the upper limit. Due to the strong SSA below $10^{12}$ Hz, we fit only the \emph{Spitzer} IR, optical, X-ray and $\gamma$-ray data.

\begin{figure}
	   \centering
		\includegraphics[width=280pt,height=200pt]{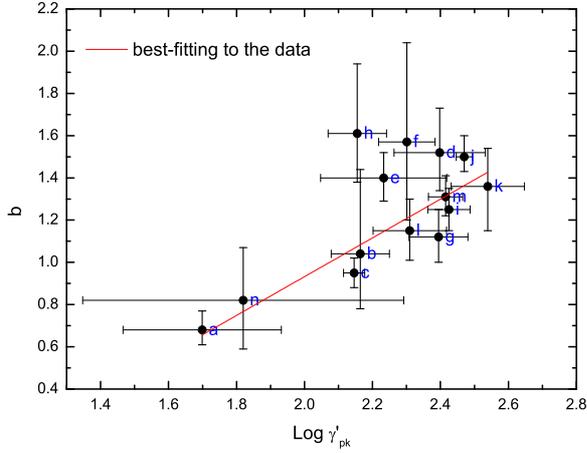}
	  \caption{Evolution of $b$ as a function of $\gamma'_{\rm pk}$. See the counterparts of the labels in Table~\ref{input}. \label{gpk-b}}
\end{figure}

\begin{figure}
	   \centering
		\includegraphics[width=280pt,height=200pt]{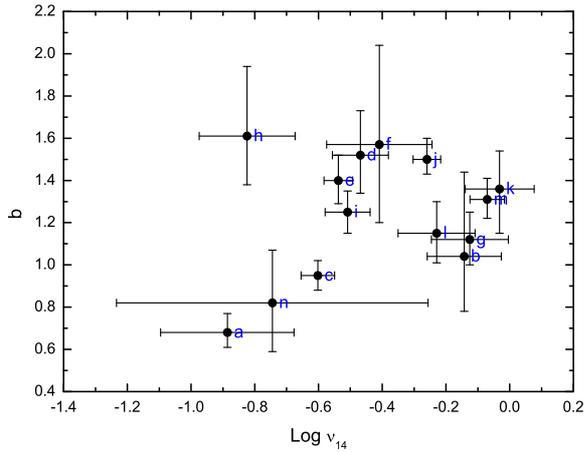}
	  \caption{Evolution of $b$ as a function of $\nu_{14}$.\label{b-nu}}
\end{figure}

\begin{figure}
	   \centering
		\includegraphics[width=280pt,height=200pt]{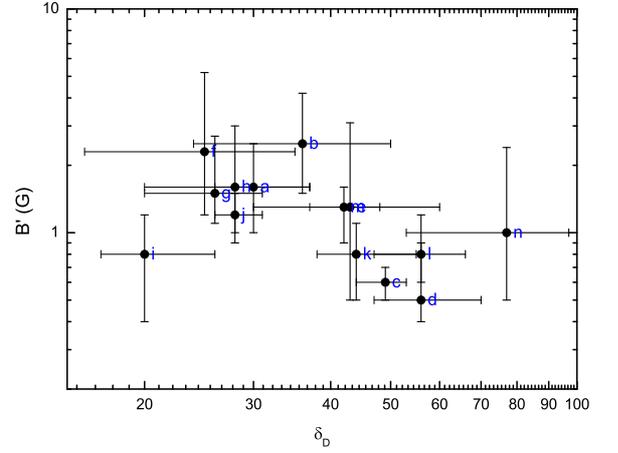}
	  \caption{Evolution of $B'$ as a function of $\delta_{\rm D}$.\label{B-D}}
\end{figure}

\begin{figure}
	   \centering
		\includegraphics[width=280pt,height=200pt]{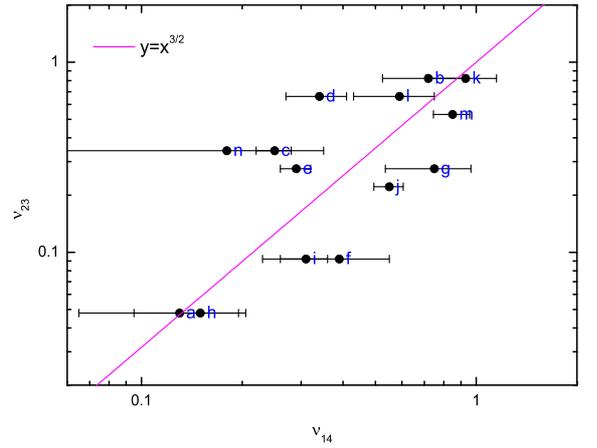}
	  \caption{Evolution of the peak frequency of the external Compton component $\nu_{23}=\nu^{\rm EC}_{\rm p}/10^{23}$ Hz as a function of $\nu_{14}$.\label{peak}}
\end{figure}

\subsection{Fitting results}

Figs.~\ref{sed1} and \ref{sed2} show the best-fit results for the 14 SEDs. The fits to the SEDs are satisfactory, though the observed data have slight excesses at highest $\gamma$-ray energies in Period A and Period C in \citet{Hayashida}.
It is noticed that the X-ray data in Period D in \citet{Hayashida15} can not be fit well with $\zeta_{\rm e}$ in [0.2, 5].
This SED can be fit well with $\zeta_{\rm e}\sim10$.

In Period A in \citet{Hayashida}, the observed X-ray emission is dominated by the EC-dust component; and in the other 13 states, the observed X-ray emission is entirely attributed to SSC.
Note that EC-BLR component
is essentially negligible in all states except Period B in \citet{Hayashida}.
Especially interesting is that the curved $\gamma$-ray spectrum during the Flare 2 in \citet{Paliya} is excellently explained by the single EC-dust component.

In Table~\ref{input} and Table~\ref{output} we give the fitting values of input and output parameters, respectively.
The mean values and the marginalized 95\% confidence intervals for free input parameters and interested output parameters are reported.
One can see that:
$b$ is in the range $\sim$[0.7, 1.6], and $\nu_{14}$ is in the range $\sim$[0.1, 0.9].
$L_{48}$ varies in the range $\sim$[0.1 - 0.6]. The emission site is required to be less than one pc from the central black hole.
The emission region is even outside the dust torus during Periods C and D in \citet{Hayashida}, and during Flare 1 and post flare in \citet{Paliya}. In the other ten states, the emission region is inside the dust torus, but is outside the BLR. In 13 flaring states, we derive similar accretion disk luminosities, i.e., $L_{\rm disk}\sim2\times10^{45}\rm\ erg\ s^{-1}$.
These values are in agreement with those obtained in earlier studies \citep[e.g.,][]{1999ApJ...521..112P}.
In the low-activity state [period H in \citet{Hayashida}], the maximum
disk luminosity allowed by the optical-UV SED is about $0.6\times10^{45}\rm\ erg\ s^{-1}$.

In Table~\ref{output}, one can see that $B'$ varies in the range $\sim$[0.2 - 2] G; $\delta_{\rm D}$ and $\gamma_{\rm pk}^{\prime}$
varies in the range $\sim$[20 - 80] and $\sim$[50 - 300], respectively.

We give the jet powers for a two-sided jet in the form of radiation ($P_{\rm r}$) and Poynting flux ($P_{\rm B}$). The relativistic
emitting electrons ($P_{\rm e}$) can be derived by the relation $P_{\rm e}=(1+\zeta_{\rm e})P_{\rm B}$, which is not listed in Table~\ref{output}.
One can see that in 8 states the radiation power is much greater than
the magnetic-field and relativistic electron power; in the other 6 states (Periods A, B, C, E in Hayashida et al. 2012, and Flare 1 as well as the post flare in Paliya et al. 2015) $P_{\rm r}$ is comparable with $P_{\rm B}(1+\zeta_{\rm e})$. 3C 279 has a black hole with mass (3 - 8)$\times10^8M_\odot$ \citep[e.g.,][]{Gu,Woo}. The Eddington luminosity is therefore in the range (4 - 10)$\times10^{46}\rm \ erg\ s^{-1}$. For low baryon loading, the total jet powers in our model are below the Eddington luminosity of 3C 279.

In Appendix~\ref{app}, we show the two-dimensional contours of the input parameters.
One can see the degeneracies between the input parameters, e.g., the degeneracy between $r$ and $\zeta_{\rm e}$.

\subsection{Correlations of model parameters}

In Fig.~\ref{gpk-b}, we show the evolution of $b$ as a function of $\gamma'_{\rm pk}$.
It is found that there is a significant correlation between $b$ and $\gamma'_{\rm pk}$, $b=0.95 \rm log\ \gamma'_{\rm pk}-0.94$, with an adjusted $R^2\approx0.8$ and a chance probability $p\approx3\times10^{-5}$.

In Fig.~\ref{b-nu}, we show the evolution of $b$ as a function of $\nu_{14}$.
There is no significant correlation between them ($R^2\approx0.4$ and $p\approx0.01$).

No clear correlation is found between $B'$ and $\delta_{\rm D}$ (see fig.~\ref{B-D}), which means that the value of $B'\cdot\delta_{\rm D}$ doest not keep $\sim$ constant during these activities.

In Fig.~\ref{peak}, we show the evolution of the peak frequency of the external Compton component ($\nu^{\rm EC}_{\rm p}=\nu_{23}\times10^{23}$ Hz) as a function of $\nu_{14}$. It seems that $\nu_{14}$ and $\nu_{23}$ follow a relation $\nu_{23}\propto\nu^{3/2}_{14}$.

\section{Summary and conclusions}

We have analysed 14 high-quality SEDs of 3C 279 in different states using a single zone leptonic model with the log-paranola EED \citep{Cerruti,Dermer14} and the MCMC technique \citep{Yan13,Yan151}. We derived the 95\% confidence intervals for the model parameters, and investigated the correlations between model parameters.

All 14 SEDs can be satisfactorily fit by the single zone leptonic model.
In 13 sates \citep[the outlier is Period B in][]{Hayashida}, the observed $\gamma$-ray emission is attributed to the EC-dust component.
The location of emission region is constrained to be outside the BLR, which is in agreement with the result derived by \citet{Nalewajko14} who used an independent approach to locate the $\gamma$-ray emission zone in the blazar jet.
Recently, \citet{Pacciani} analysed the SEDs in $\gamma$-ray
flares of 10 FSRQs, and also found that the $\gamma$-ray emission takes place outside the BLR.
In the Thomson regime, the peak frequency of the EC component will strongly correlate with the synchrotron peak frequency if $\zeta_{\rm e}\sim1$ \citep{Dermer14}. In our analysis, the large dispersion of the correlation between the two peak frequencies (Fig.~\ref{peak}) is due to the uncertainty on $\zeta_{\rm e}$.

It is worth pointing out that all SEDs except the SED in Period D in \citet{Hayashida} can be fitted well in near-equipartition conditions ($\zeta_{\rm e}\sim$ a few). \citet{Sironi2} recently found that in the magnetic reconnection powering jet emission scenario the emitting region is characterized by a rough energy equipartition between magnetic fields and radiating particles.

It is interesting to note that the curved $\gamma$-ray spectrum in Flare 2 in \citet{Paliya} can be excellently explained by the EC-dust component alone (see Fig.~\ref{sed1}).
The curvature/break \citep{Harris} in the $\gamma$-ray spectrum was discovered in the \emph{Fermi}-LAT (Large Area Telescope) observations for 3C 454.3 \citep{Abdo09}. Several approaches have been proposed to explain this feature,
including the combination of Compton-scattered disk and BLR radiations \citep{Finke10}, and IC scattering BLR radiation
in Klein-Nihisna regime \citep{Cerruti}. \citet{Paliya} showed that
the superposition of EC-BLR component and EC-dust component can account for the curved GeV spectrum of 3C 279.
\citet{Paliya1} has very recently found that the $\gamma$-ray spectrum of 3C 279 during an exception $\gamma$-ray outburst in 2015 June also shows a clear signature of break/curvature.

More importantly, we found that there is a significant positive correlation between
the curvature parameter $b$ of EED and the electron peak energy $\gamma'_{\rm pk}$.
\citet{Tramacere} showed that when cooling becomes relevant, both $b$ and $\gamma'_{\rm pk}$ increase in the case of $q=2$ as the EED approaches the equilibrium.
The positive correlation between $b$ and $\gamma'_{\rm pk}$ found here is consistent with
the prediction in the cooling-relevant scenario in the case of $q=2$.
A correlation between $b$ and $\nu_{14}$ ($\nu_{14}\propto B'\delta_{\rm D}\gamma^{\prime 2}_{\rm pk}$) is predicted if $B'\cdot\delta_{\rm D}$ roughly remains constant from
state to state. However, we find no significant correlation between $b$ and $\nu_{14}$ (Fig.~\ref{b-nu}) due to varied $B'\cdot\delta_{\rm D}$ from state to state (Fig.~\ref{B-D}).
These situations are different from those \citep[e.g.,][]{Tramacere07,Tramacere09} for HSP BL Lacs.

\cite{Chen} fitted the synchrotron bumps of a sample of \emph{Fermi} blazars using a log-parabola function, and claimed that the inverse correlation between $b_{\rm syn}$ and $\nu_{\rm s}$ holds for the \emph{Fermi} blazars half of which are FSRQs, which is inconsistent with our result.

Note that the increase in $b$ only happens in the case of the hard-sphere scenario, i.e., the turbulence spectral index $q=2$ \citep{Tramacere}. Therefore, our results support the hard-sphere approximation, in agreement with the magnetohydrodynamic simulations in \citet{Brandenburg}. In the case of $q=3/2$, EED approaches the equilibrium with a stable $b$.

In addition to the decrease of $b$ in acceleration-dominated scenario and the increase in cooling-relevant scenario, \citet{Tramacere} showed that when EED reaches the equilibrium, the curvature of EED reaches a stable value.
They suggested that such a limit value could be found in cooling-dominated flares in HSP BL Lacs.
\citet{Yan13} modeled two SEDs of Mrk 421, respectively, in low state \citep{Abdo11} and in giant flare \citep{Shukla}, using the SSC model with a power-law log-parabola (PLLP) EED. We found that the curvature of EED increased with the electron peak energy and $\nu_{14}$, which hints that the EED during the
giant flare is at the equilibrium or very close.
While radiative cooling in FSRQs is more severe, and EED is easy to achieve equilibrium.
It seems that the limit value of $b$ for 3C 279 is $\sim$1.6 (see Fig.~\ref{gpk-b}).

Recently, \citet{Asano15} successfully reproduced the SED in Period D in \citet{Hayashida} by using a steady stochastic acceleration model with $q=2$. We found $b\sim1.5\pm0.2$ during this period, which implies the EED in this state being at the equilibrium too.
The result in \citet{Asano15} supports our above discussions. In the magnetic reconnection scenario, particles can also be efficiently accelerated to develop nonthermal distributions \citep{zh,Guo2014,Sironi,Werner}. Three-dimensional numerical simulations show that the magnetic reconnection generates magnetic turbulence \citep{Guo2015}, which may further drive strong stochastic acceleration.
It may be interesting to systematically reproduce the 14 SEDs in the stochastic acceleration model \citep[e.g.,][]{Yan12,ChenAS,Asano15} to further check our discussions.

In summary, we have analysed 14 high-quality and simultaneous SEDs of 3C 279 \citep{Hayashida,Hayashida15,Paliya}, using the approach in \citet{Yan151}.
We found that there is a clear positive correlation between $b$ and $\gamma'_{\rm pk}$.
Based on the calculations in \citet{Tramacere}, this correlation may hint that radiative cooling of electron is relevant, and the EEDs are close to the equilibrium. No clear correlation between $b$ and $\nu_{14}$ is found, because of the varied value of ($B'\cdot\delta_{\rm D}$) from state to state.
Other results worthy of remark are: (1) the allowed minimum variability timescale during the extremely bright $\gamma$-ray flare in 2014 March-April can be as short as $\sim3000\ $s; (2) the alone EC-dust component can account for the curvature in the $\gamma$-ray spectrum of 3C 279.

\section*{Acknowledgments}
We thank Chuck Dermer for his valuable suggestions and questions, and Fan Guo for discussions.
We thank the anonymous referee for helpful suggestions.
We are grateful to Krzysztof Nalewajko and Vaidehi Paliya for providing us the data of 3C 279.
DHY thanks Qiang Yuan for help on MCMC technique.
This work is partially supported by the National Natural Science Foundation of China (NSFC 11433004) and Top Talents Program of Yunnan Province, China. DHY acknowledges partial funding support by China Postdoctoral Science Foundation under grant no. 2015M570152, and by the National Natural Science Foundation of China (NSFC) under grant no. 11573026.
 SNZ acknowledges partial funding support by 973 Program of China under grant 2014CB845802, by the National Natural Science Foundation of China (NSFC) under grant nos. 11133002 and 11373036, by the Qianren start-up grant 292012312D1117210, and by the Strategic Priority Research Program ``The Emergence of Cosmological Structures'' of the Chinese Academy of Sciences (CAS) under grant no. XDB09000000.

\bibliography{refernces}

\appendix
\section{Two-dimensional contours of the input parameters}
\label{app}

\begin{figure*}
	   \centering
		\includegraphics[width=250pt,height=165pt]{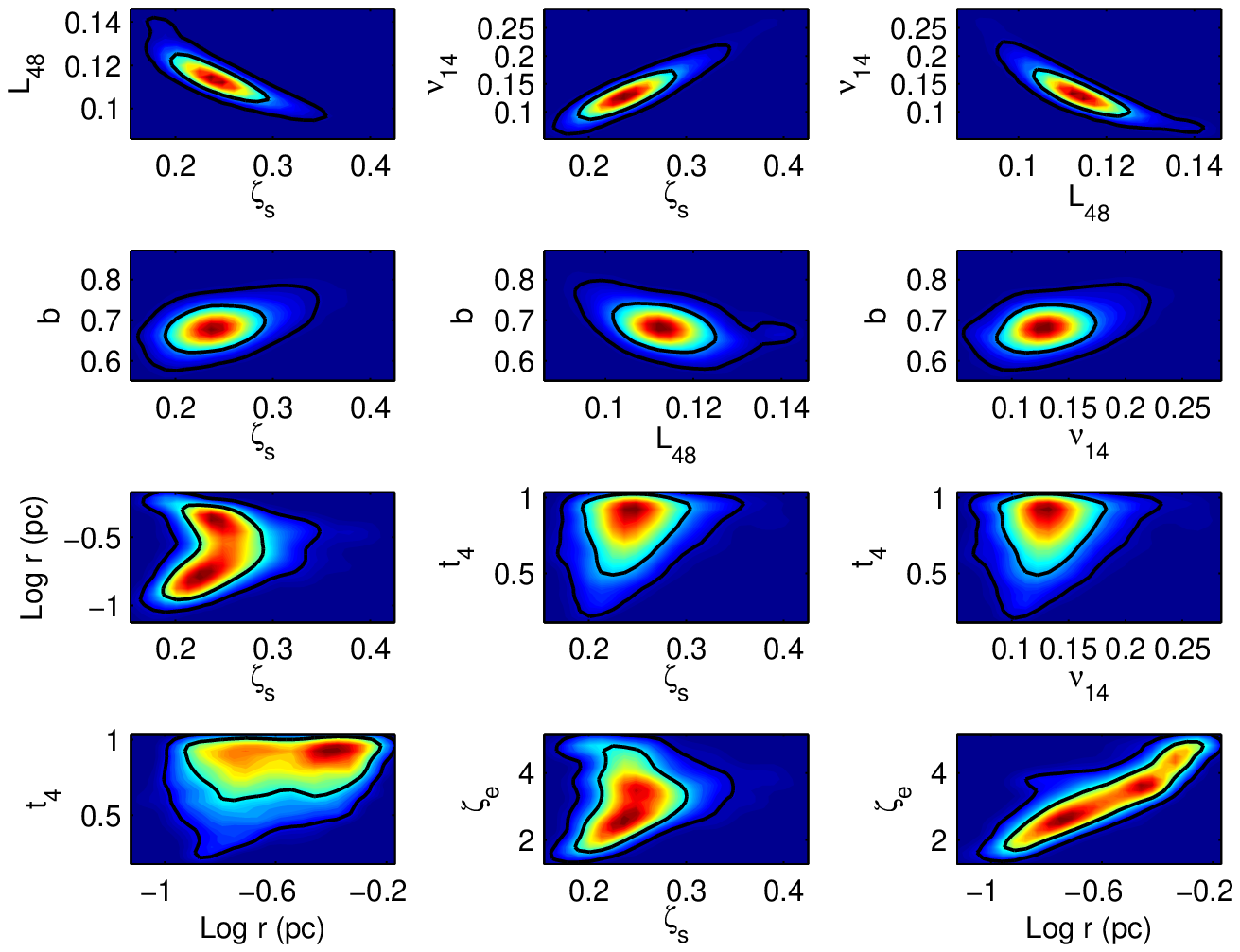}
        \includegraphics[width=250pt,height=165pt]{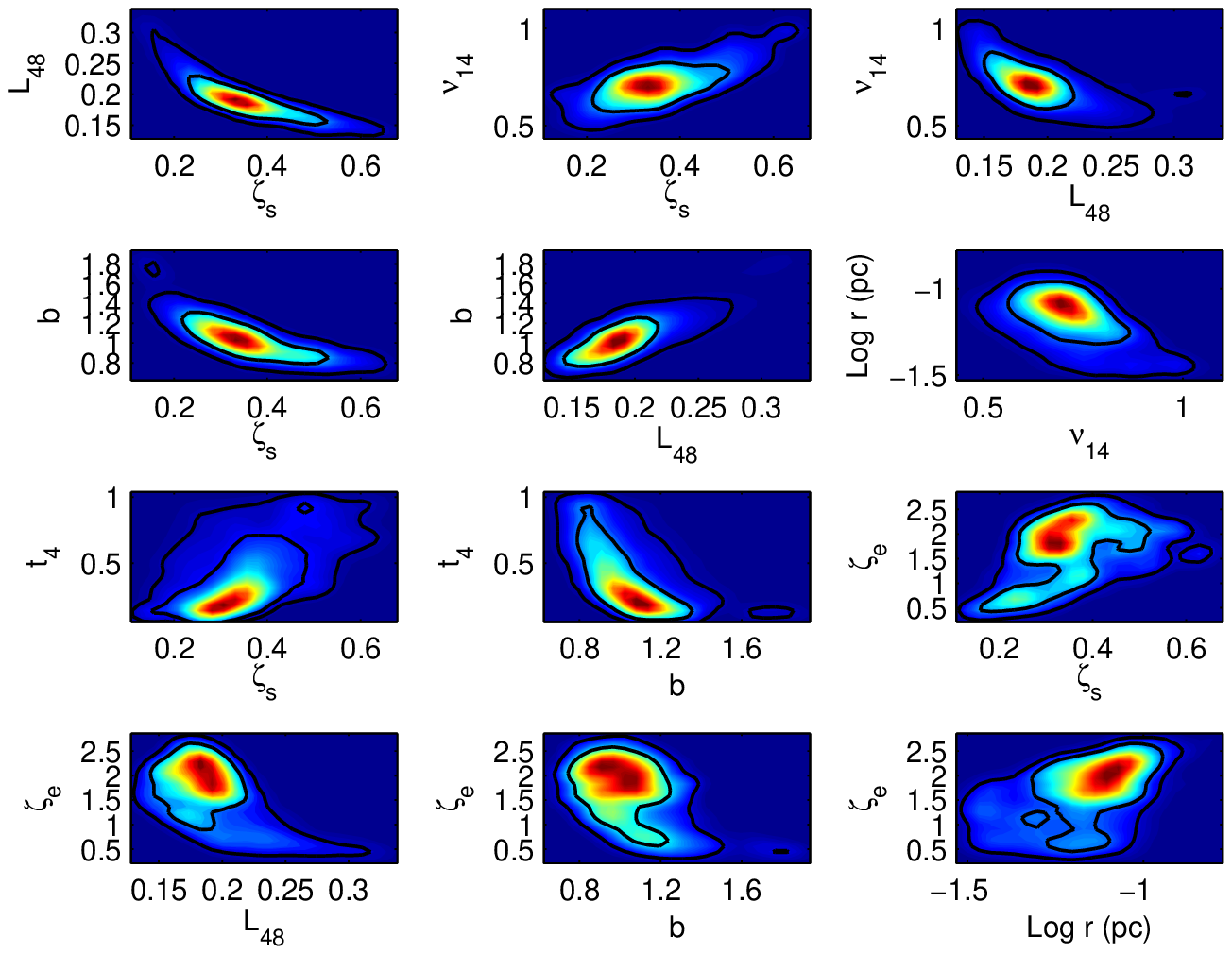}
        \includegraphics[width=250pt,height=165pt]{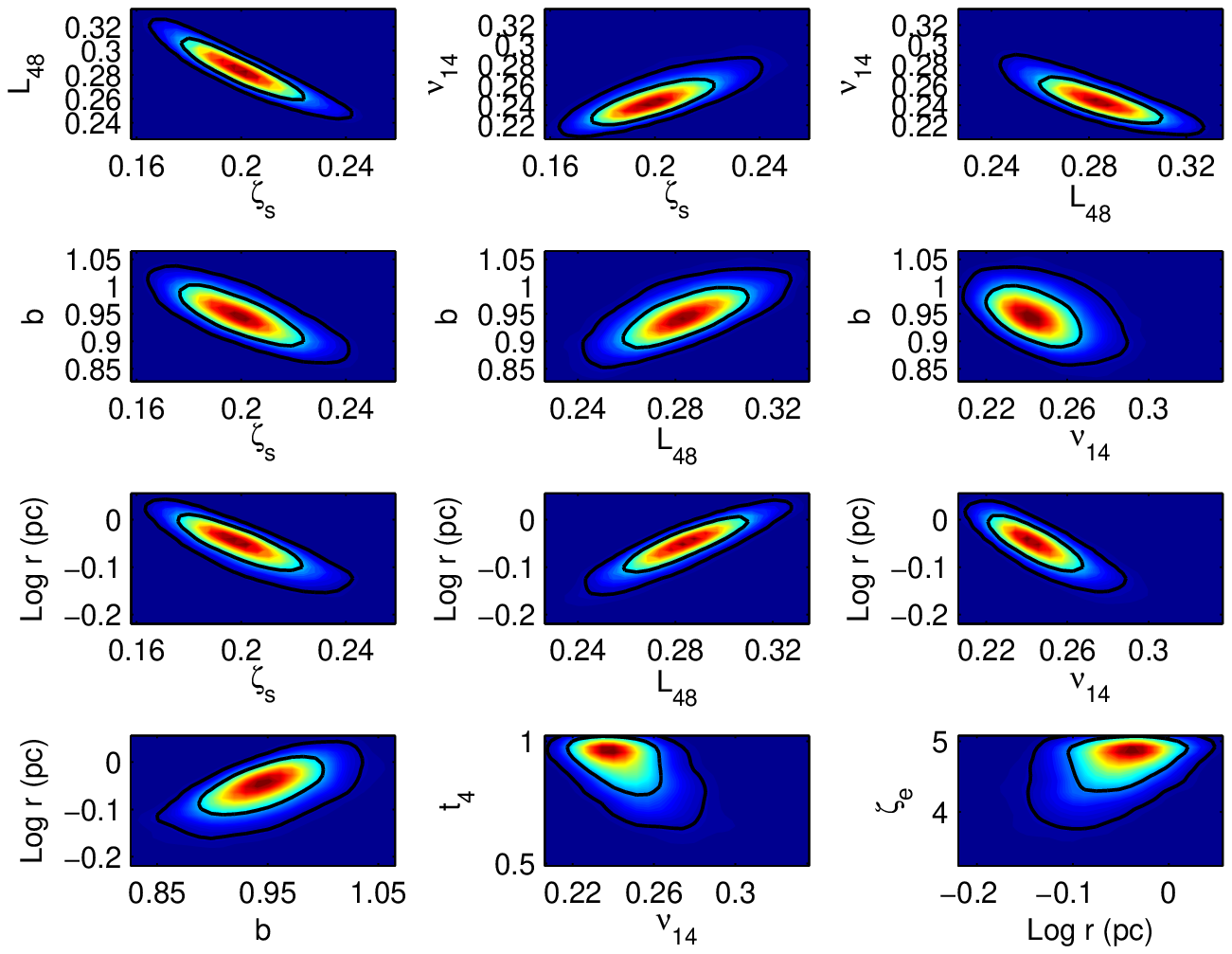}
        \includegraphics[width=250pt,height=165pt]{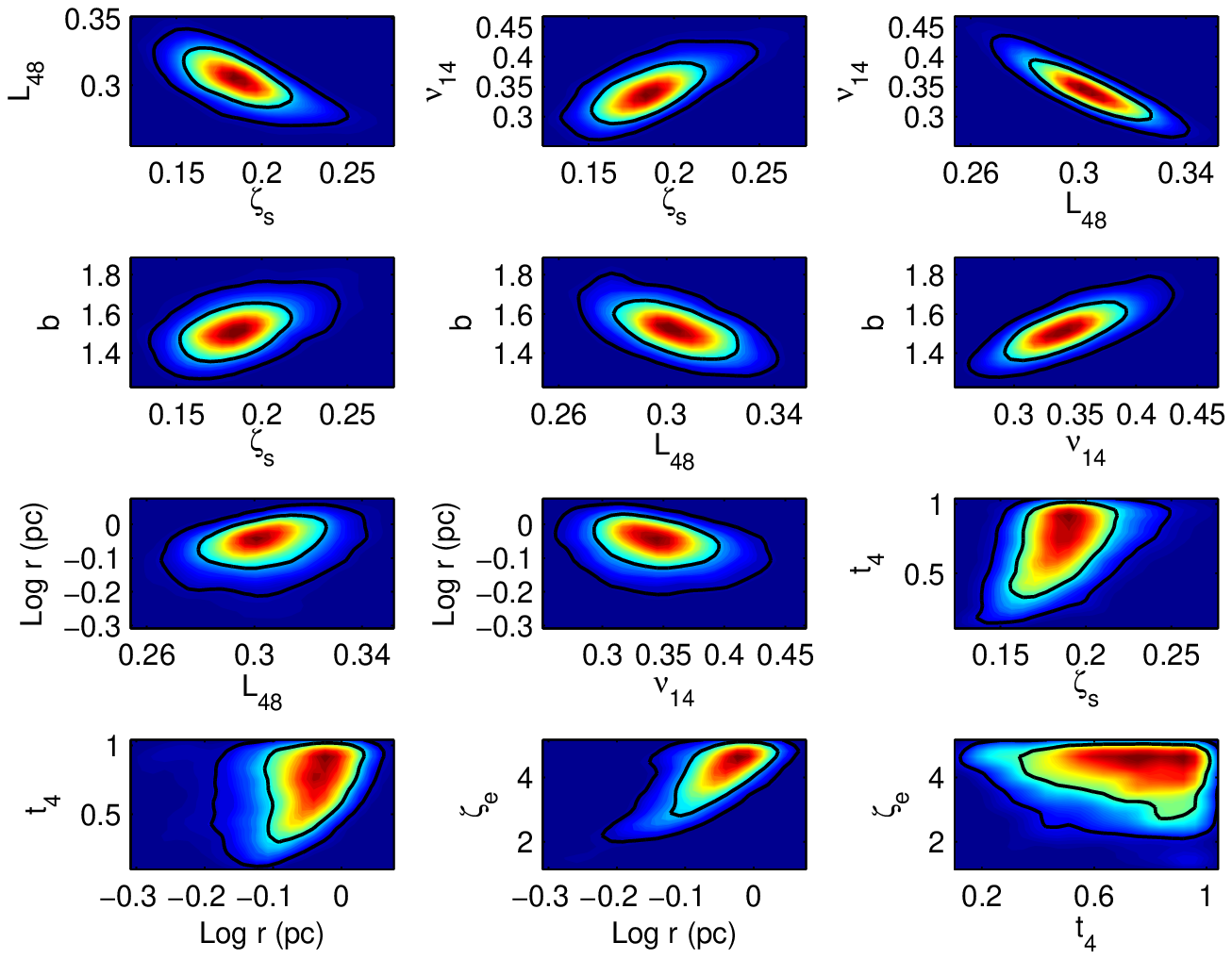}
        \includegraphics[width=250pt,height=165pt]{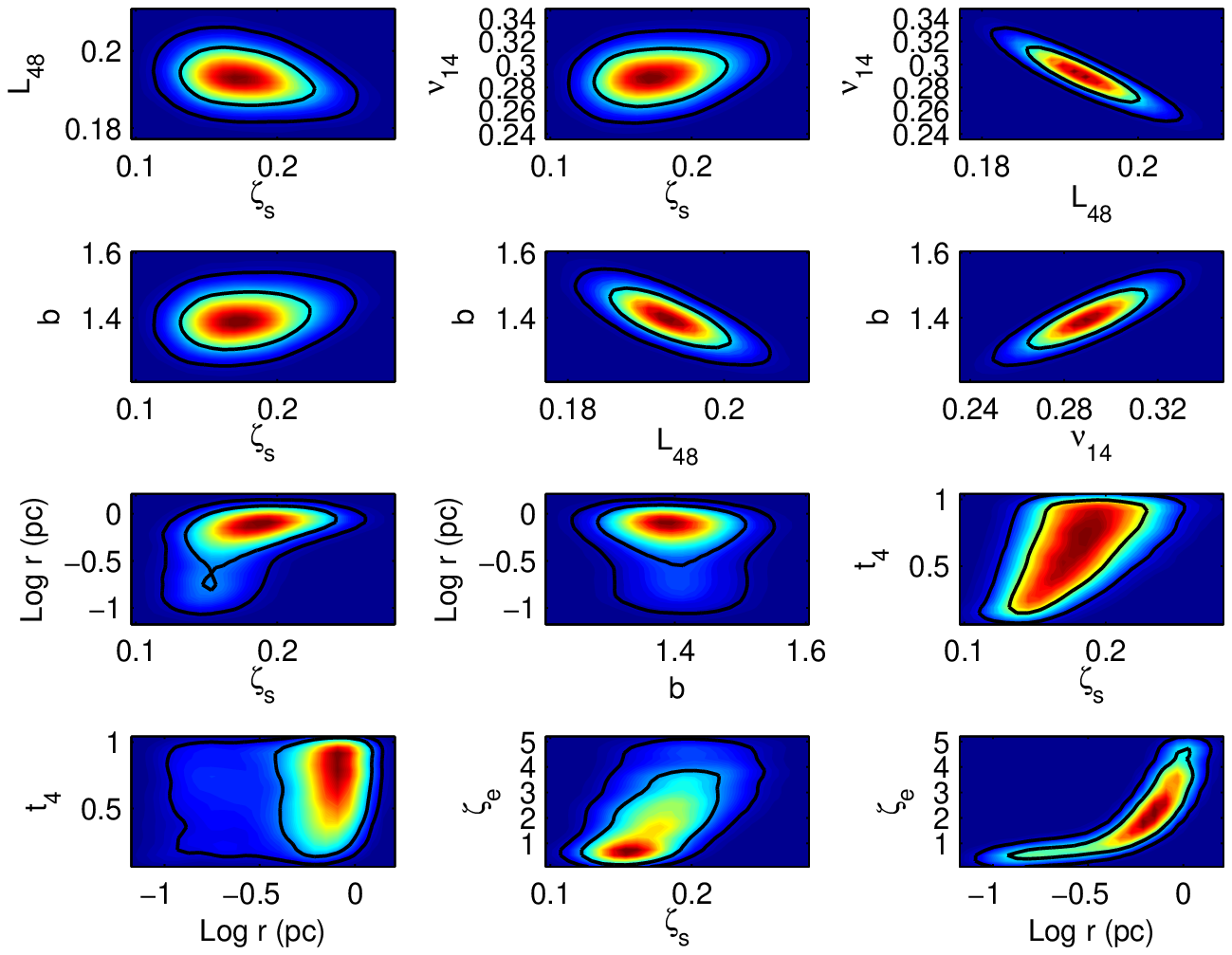}
        \includegraphics[width=250pt,height=165pt]{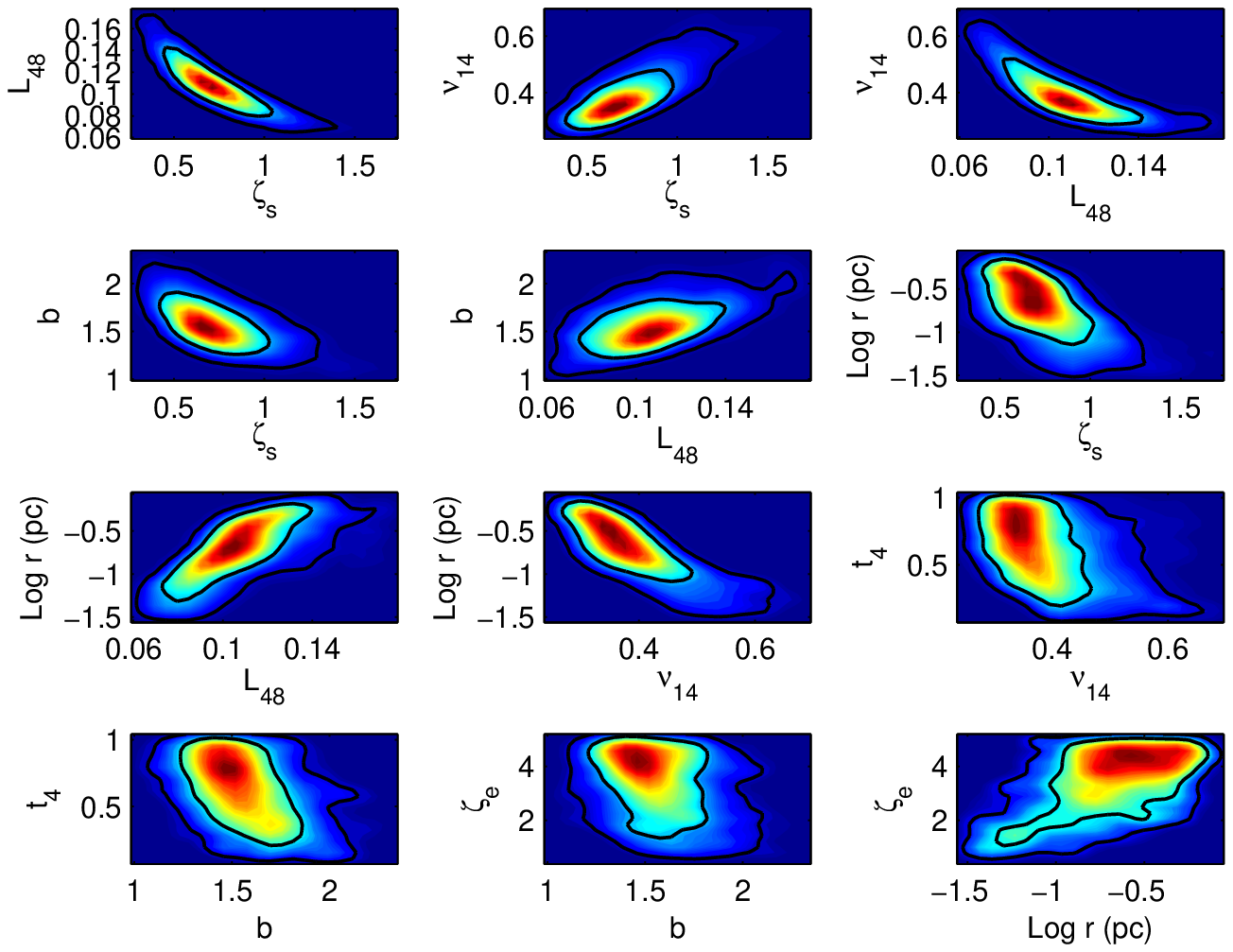}
        \includegraphics[width=250pt,height=165pt]{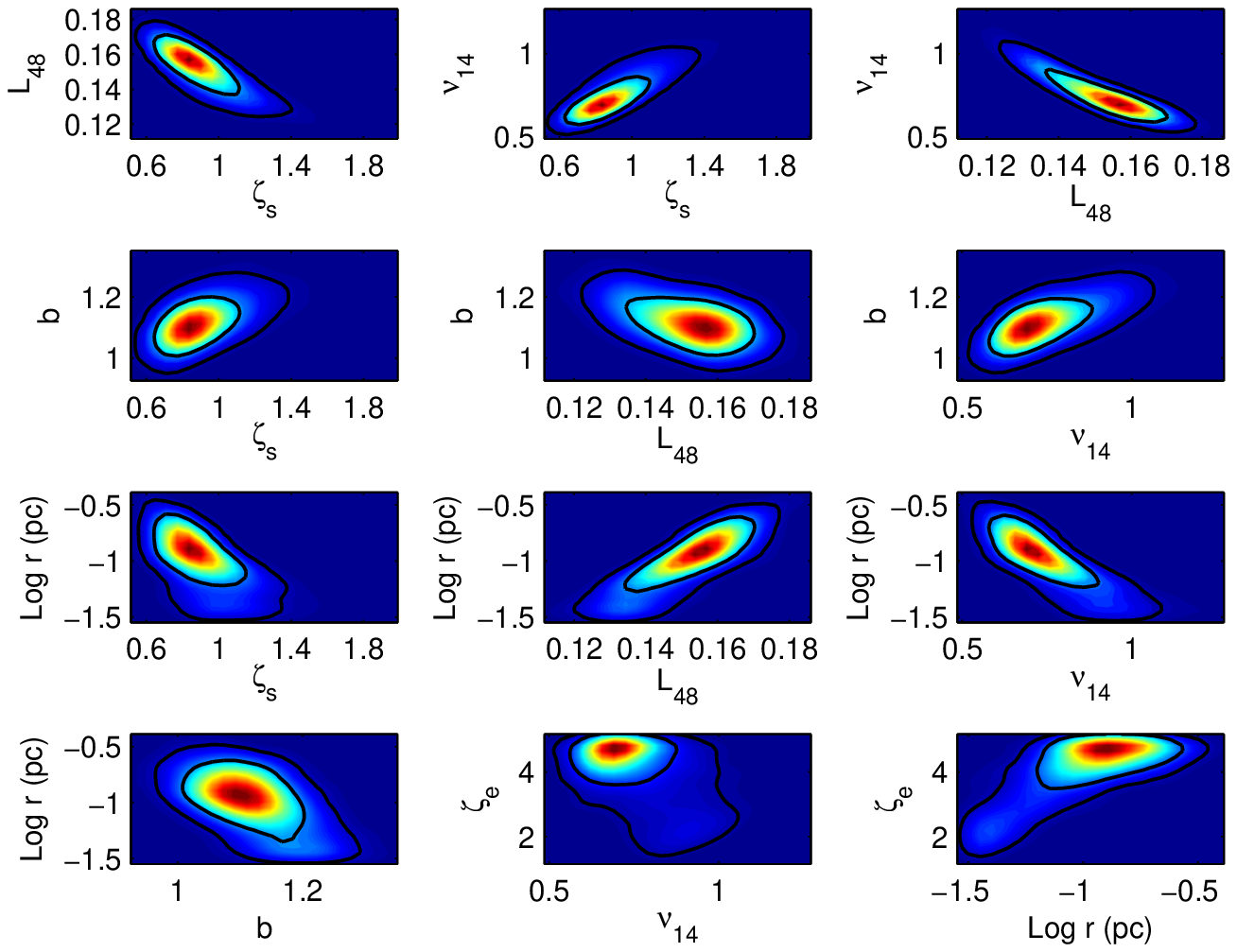}
        \includegraphics[width=250pt,height=165pt]{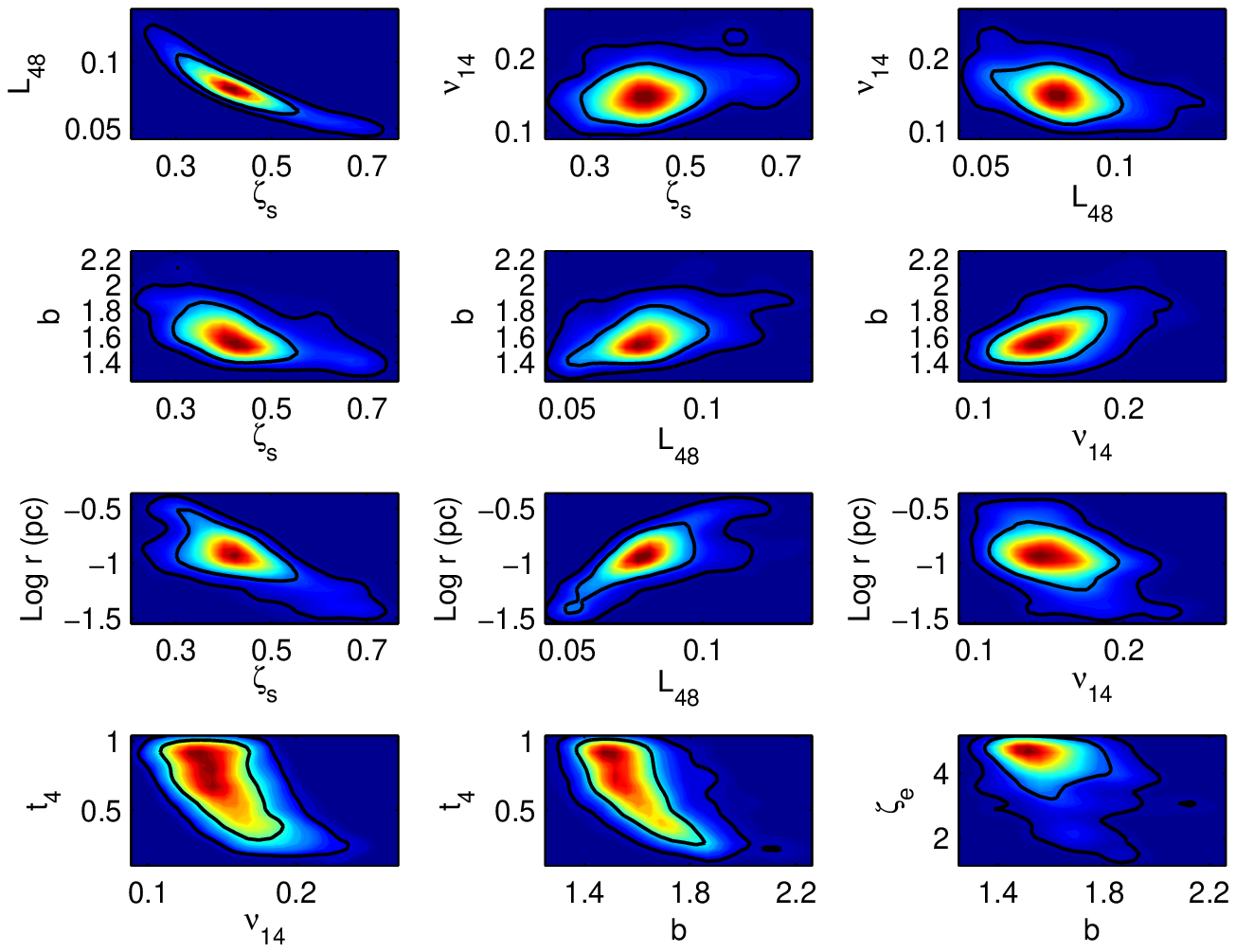}
	  \caption{Two-dimensional contours of free input parameters [the regions enclosing the 68 per cent (95 per cent) confidence level are shown] derived in the fitting to the 8 SEDs of 3C 279 in \citet{Hayashida}. The panels correspond to those in Fig.~\ref{sed1}. \label{2D1}}
\end{figure*}

\begin{figure*}
	   \centering
		\includegraphics[width=250pt,height=165pt]{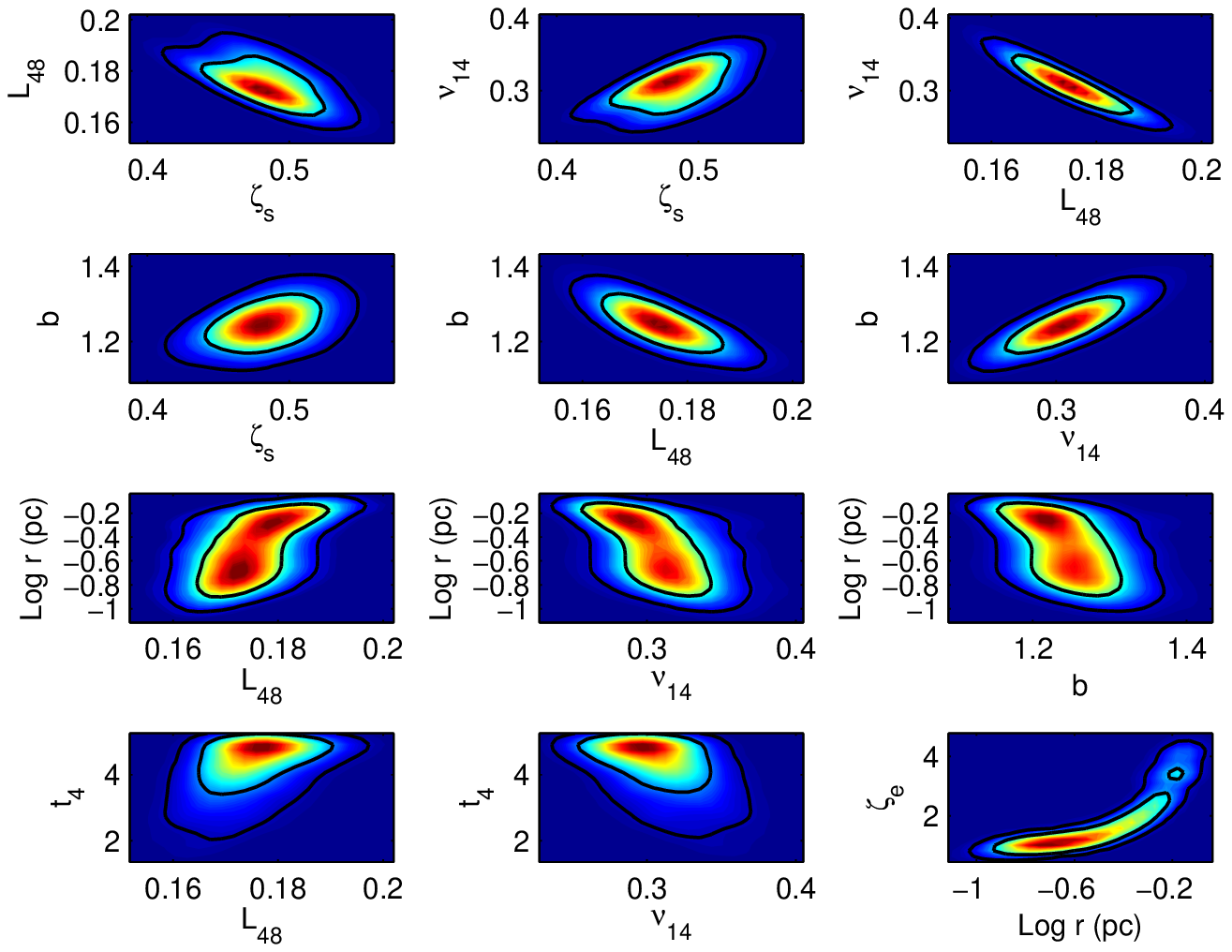}
        \includegraphics[width=250pt,height=165pt]{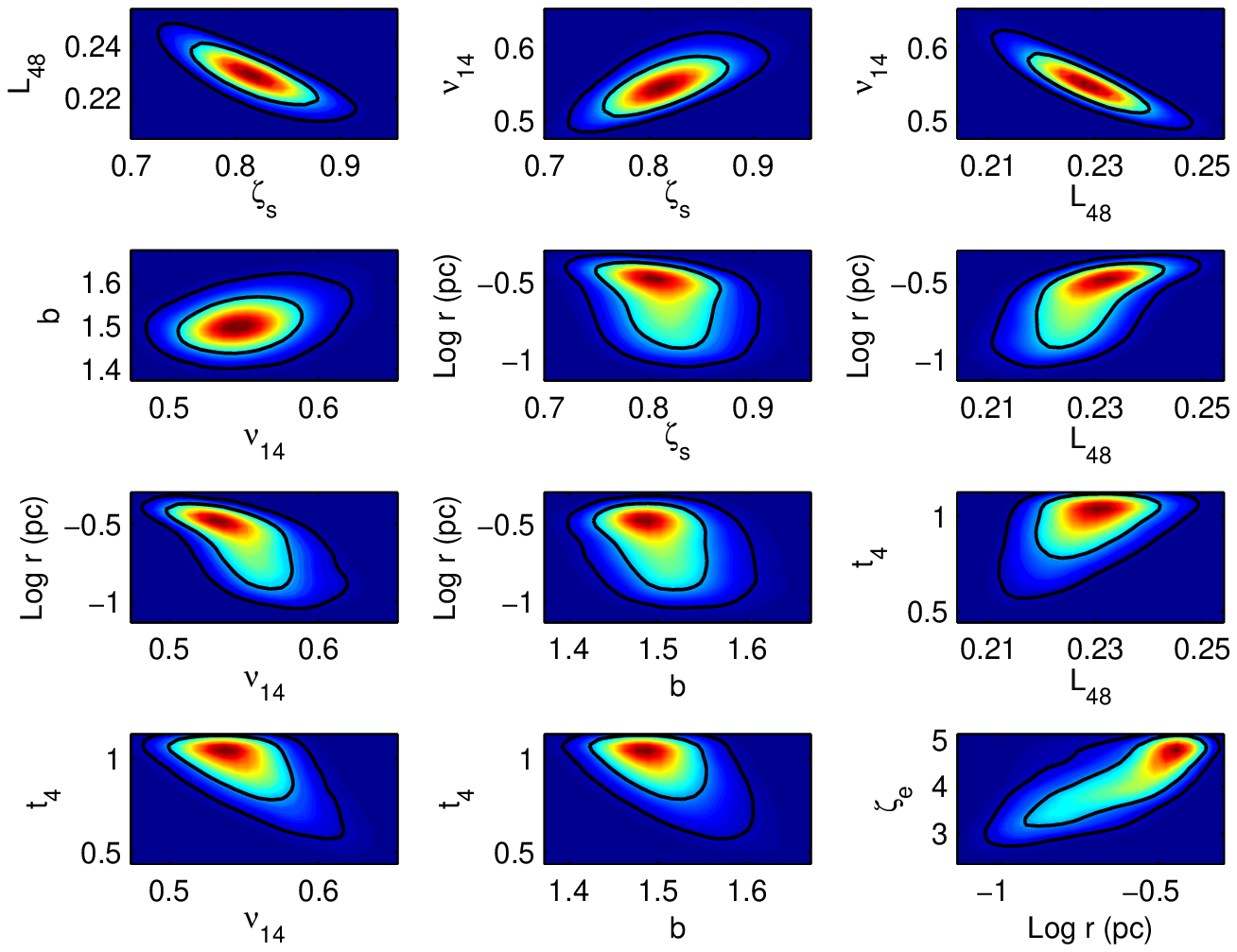}
        \includegraphics[width=250pt,height=165pt]{12D_in.eps}
        \includegraphics[width=250pt,height=165pt]{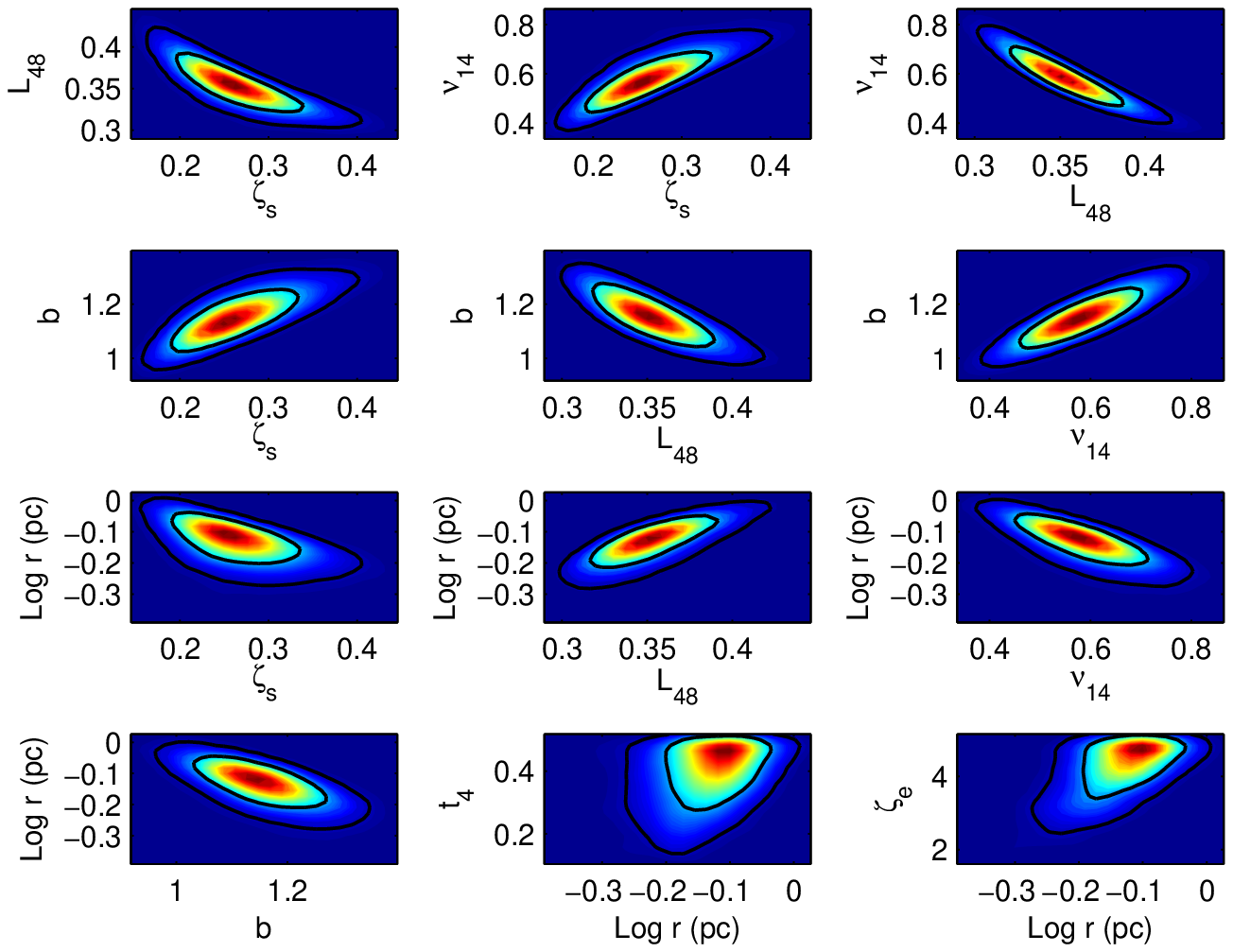}
        \includegraphics[width=250pt,height=165pt]{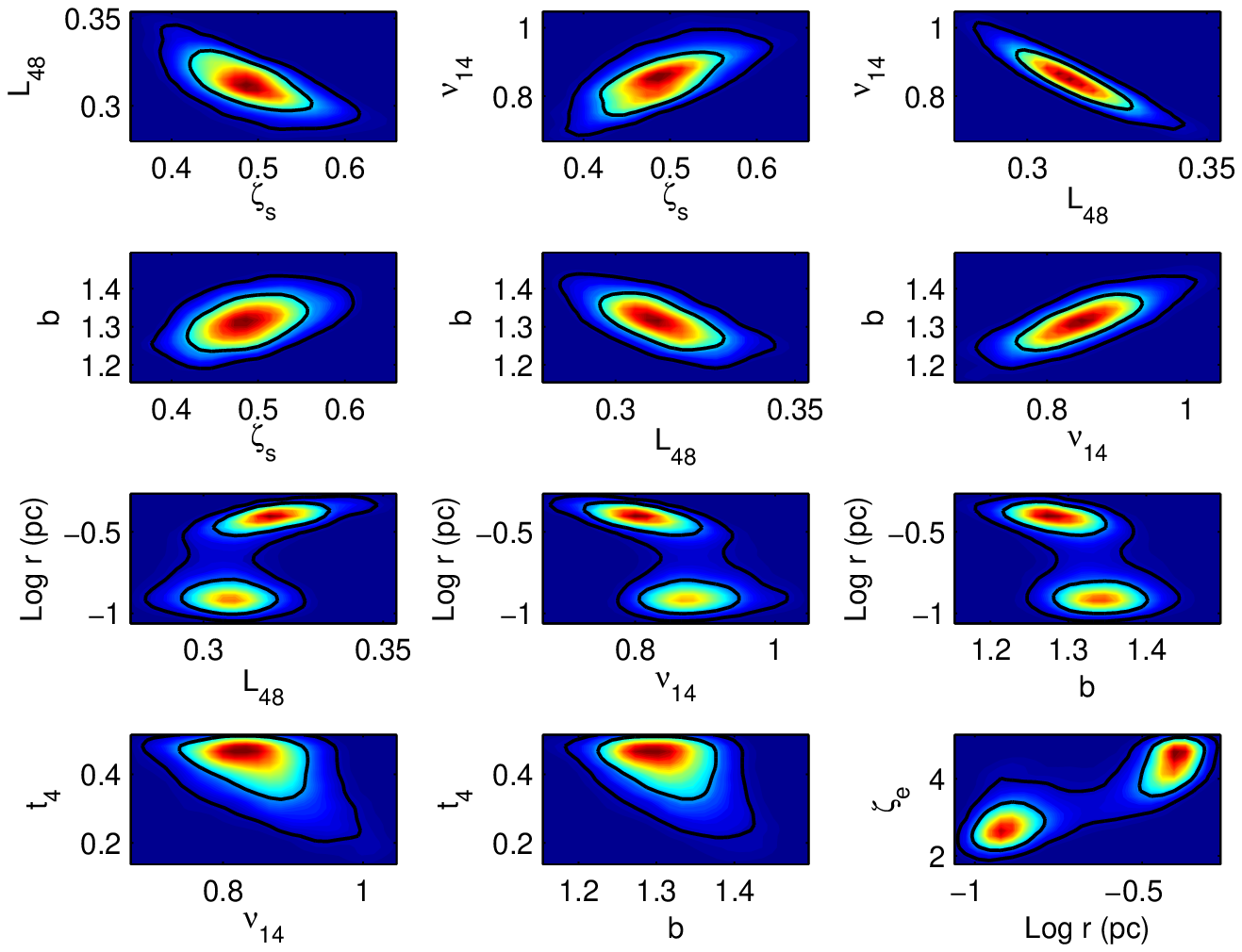}
        \includegraphics[width=250pt,height=165pt]{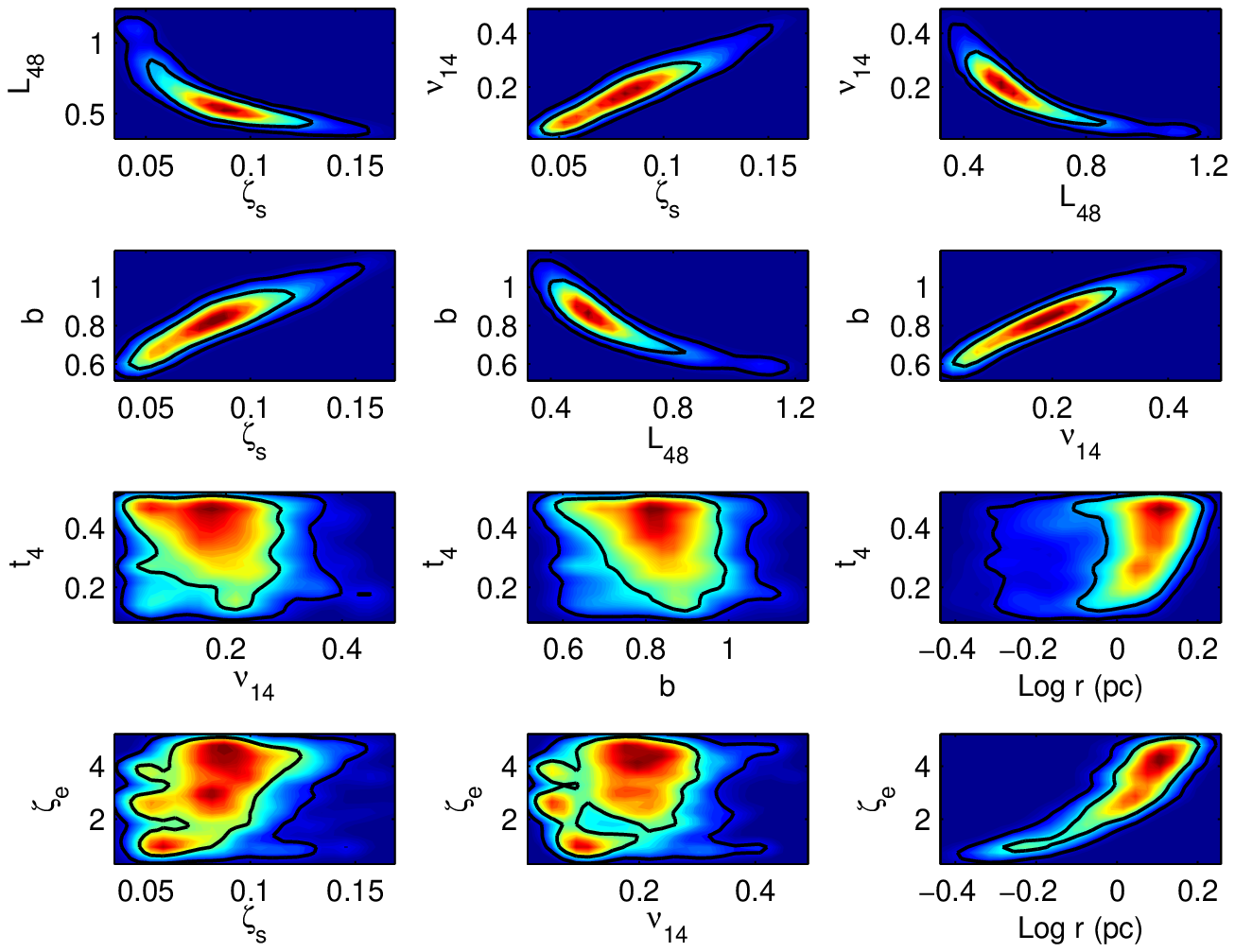}
	  \caption{Two-dimensional contours of free input parameters [the regions enclosing the 68 per cent (95 per cent) confidence level are shown] derived in the fitting to the 3 SEDs of 3C 279 in \citet[][]{Hayashida15}, and to the 3 SEDs in \citet{Paliya}. The panels correspond to those in Fig.~\ref{sed2}. \label{2D2}}
\end{figure*}

We give the two-dimensional contours of the input parameters derived in our fittings in Figs.~\ref{2D1} and \ref{2D2}.
One can look at the degeneracies between the input parameters.

\bsp	
\label{lastpage}
\end{document}